# Single-Crystal Ir$_3$Te$_8$: Superconductivity Emerging from a Diamagnetic, Non-Fermi-Liquid Normal State

L. Li[1], T. F. Qi[1], L. S. Lin[2], X. X. Wu[3], X.T. Zhang[3], K. Butrouna[1], V. S. Cao[1#], Y. H. Zhang[2], Jiangping Hu[3,4], P. Schlottmann[5], L. E. De Long[1] and G. Cao[1]

[1] Department of Physics and Astronomy and Center for Advanced Materials, University of Kentucky, Lexington, KY 40506, USA
[2] High Magnetic Field Laboratory, Chinese Academy of Sciences, Hefei 230031, China
[3] Beijing National Laboratory for Condensed Matter Physics, Institute of Physics, Chinese Academy of Sciences, Beijing, China
[4] Department of Physics, Purdue University, West Lafayette, IN 47907, USA
[5] Physics Department, Florida State University, Tallahassee, FL 32306, USA

**Abstract**

We observe a highly unusual combination of normal and superconducting state properties without any signature of strong spin fluctuations in single-crystal Ir$_3$Te$_8$. A dominant linear temperature dependence of electrical resistivity extends from 20 K to 700 K, in spite of a strongly first-order phase transition from cubic to rhombohedral lattice symmetry below $T_S$ = 350 K. Although the electronic heat capacity coefficient is substantial (11 mJ/mole-K$^2$), a highly diamagnetic rather than Pauli paramagnetic normal state yields to superconductivity below a critical temperature $T_C$ = 1.8 K. Electronic structure calculations indicate two bands cross the Fermi level; one band is quasi-two-dimensional, and may be responsible for the observed diamagnetism and structural transition.

Ir compounds have recently emerged as a fertile ground for discoveries of new physics driven by large spin-orbit interactions (SOI) typical of 5d electronic states. The close competition between the SOI, crystalline electric fields, and the Coulomb interaction U stabilizes novel ground states and phenomena [1-6]. These discoveries have stimulated a variety of theoretical predictions [7-13] that await experimental confirmation, including topologically driven semimetallic [12] or superconducting [13] states.

In this Letter, we report a highly unusual combination of phenomena in single-crystal $Ir_3Te_8$: In spite of a substantial electronic heat capacity coefficient $\gamma = 11$ mJ/mole-K$^2$, the normal state of $Ir_3Te_8$ exhibits strong diamagnetism rather than Pauli paramagnetism. Moreover, the electrical resistivity does not exhibit the usual T$^2$ Fermi liquid term, but remains strongly linear in temperature from 20 K to 700 K, and remains large ($\approx$ 0.46 m$\Omega$-cm) below 4 K. The linear-T resistivity persists through a first-order phase transition at $T_S$ = 350 K which separates a high-temperature cubic phase from a low-temperature rhombohedral phase. Although the unusual properties of $Ir_3Te_8$ suggest a non-Fermi liquid state, superconductivity emerges below a critical temperature $T_C$ = 1.8 K. Most structural and physical properties reported here are unique to single-crystal $Ir_3Te_8$ and have not been observed in polycrystalline samples [20].

Superconductivity has recently been found in Ir-based chalcogenides, namely $CdI_2$-type $Ir_{1-x}M_xTe_2$ (M = Pd and Pt) [14-17], and $Ir_xSe_2$ (x > 0.75) pyrites [18]. The superconducting $Ir_{1-x}M_xTe_2$ is a derivative of layered $IrTe_2$, which exhibits a structural transition near 260 K, from a rhombohedral, to a low-temperature, monoclinic structure. This transition was attributed to a possible charge density wave (CDW) [14], although no CDW gap has yet been detected [16, 17]. Nevertheless, superconducting transitions with $T_C$ as high as 3 K emerge with the suppression of the structural transition in doped $Ir_{1-x}M_xTe_2$ [14], similar to classic CDW compounds such as $NbSe_2$ and $(Nb_{1-x}Ta_x)Se_3$ [18, 19]. Higher $T_C$'s of up to 6.4 K were found in a polycrystalline $Ir_{0.91}Se_2$ pyrite [20], which is a nonstoichiometric variant of a parent $Ir_{0.75}Se_2$ (or $Ir_3Se_8$) compound of rhombohedral space group *R-3*, and is proximate to a metal-insulator transition. We note that no high-temperature phase transitions similar to that found in layered $IrTe_2$ are reported for relatively high-$T_C$ $Ir_xSe_2$ compositions [20]. Superconductivity in these Ir-based chalcogenides is evidently not as sensitive to crystal structure as it is for many other transition metal materials, for which narrow peaks in the electronic density of states promote $T_C$'s that range from 10 to 20 K [21]. Moreover, these chalcogenides sharply contrast with known Ir oxides, where a strong SOI generally drives narrow-gap insulating states [1-2, 22-26].

Single crystals of $Ir_3Te_8$ were grown by a slow-cooling technique using excess Te as flux. Stoichiometric quantities of the elements were ground thoroughly and sealed in an evacuated quartz tube, which was slowly heated up to 1000 ºC and held at that temperature for 7 days. The synthesized polycrystalline $Ir_3Te_8$ was then mixed with an appropriate amount of Te powder and sealed under vacuum in a small quartz tube, which was subsequently put in a larger quartz tube that was then evacuated and sealed. The mixture was heated up to 1050 ºC, where it was maintained for over 48 hours, followed by slow-cooling to 700 ºC. The average size of the single crystals was 1 x 1 x 1 $mm^3$, as shown in **Fig. 1**. Measurements of magnetization M(T,H), heat capacity C(T) and electrical resistivity ρ(T) were performed over the temperature interval 0.5 K < T < 700 K, using either a Quantum Design (QD) Physical Property Measurement System, or a QD Magnetic Property Measurement System equipped with a Linear Research Model 700 AC bridge. The high temperature ρ(T) was measured using a Displex closed-cycle cryostat capable of continuous temperature ramping from 9 K to 900 K.

Density functional theory (DFT) calculations were carried out using the projector augmented wave (PAW) method encoded in the Vienna *ab initio* simulation package (VASP) **[27-29]**, and employed the generalized-gradient approximation (GGA) for the exchange correlation functional **[30]**. The SOI of the valence electrons was included using the second-variation method for the scalar-relativistic eigenfunctions of the valence states **[31]**. The plane-wave basis set cutoff was set at 500 eV. The Brillouin zone was sampled using the Monkhorst-Pack scheme **[32]** with (7×7×7) **k**-points for the $Ir_3Te_8$ and $Ir_4Te_8$ primitive cells. The lattice parameters used were a = 6.4024 Å and α = 90.017° (measured at 90 K; see **Table 1**). The atomic positions were fully relaxed and forces were minimized to less than 0.01 eV/Å.

The crystal structure of a small single crystal was determined using Mo *Kα* radiation and a Nonius Kappa CCD single-crystal diffractometer at the temperatures 90 K, 250 K, 295 K, 350 K and 390 K. The structures were refined using the SHELX-97 programs **[33-34].** Crystal composition was examined by energy-dispersive X-ray (EDX) spectroscopy using a Hitachi/Oxford SwiftED 3000. The *R*- and $R_W$-factors are low, 0.025 and 0.058, respectively, and the mosaicity was also small, suggesting well-ordered crystals, as shown in **Figs. 1c** and **1d**. The Ir site occupancy was freely varied and found to be 80%. A superlattice was evident in all X-ray diffraction data, suggesting that the Ir vacancies may order in our crystals (see **Fig. 1b**). Given the possible existence of inhomogeneities in these materials **[15]**, the structural and physical properties of a number of $Ir_3Te_8$ crystals were examined, and we found no discernible discrepancies between data for all measured crystals.

The room-temperature crystal structure of $Ir_3Te_8$ was initially reported to be cubic with space group *Pa3* [35], but was subsequently described as rhombohedral with space group *R-3* [36]. Our single-crystal X-ray refinements show $Ir_3Te_8$ undergoes a structural phase transition near $T_S$ = 350 K, changing from a high-temperature cubic lattice with space group *Pa3* (No. 205) to a low-temperature rhombohedral lattice with space group *R-3* (No. 148) (see **Table 1**). The difference between the two structures can be defined by the difference $\Delta\theta$ ($\equiv \theta_1-\theta_2$) between two bond angles $\theta_1$ (Ir1-Te1-Ir2) and $\theta_2$ (Ir2-Te2-Ir1); i.e., $\Delta\theta$ is zero for $T \geq T_S$ = 350 K when the lattice is cubic, and finite for $T < T_S$ where the lattice symmetry is reduced to rhombohedral (see **Figs. 1a** and **1b**). The structural transition is subtle, but causes strong anomalies in the transport and magnetic properties, as discussed below.

**Table 1:** *Lattice Parameters for $Ir_3Te_8$*

| Temperature (K) | 90 | 250 | 390 |
|---|---|---|---|
| *a* (Å) | 6.4024 | 6.4081 | 6.4152 |
| $\alpha$(°) | 90.017 | 90.017 | 90 |
| Structure (Space group) | Rhombohedral (*R-3*) | Rhombohedral (*R-3*) | Cubic (*Pa3*) |

Our heat capacity data C(T) for 1.8 K $\leq$ T < 10 K yield a Debye temperature $\theta_D$ = 246 K, and an electronic coefficient $\gamma$ = 11 mJ/mole K$^2$. These values are quite similar to those for $Ir_xSe_2$ [20]. Single-crystal $Ir_3Te_8$ is distinctly metallic (although highly resistive) throughout a wide temperature range (see **Fig. 2a**), which sharply contrasts the insulating behavior of $Ir_3Se_8$ [20]. The **a**-axis electrical resistivity $\rho_a$(T) is interrupted by a strong first-order anomaly with hysteresis in the vicinity of $T_S$ = 350 K, which is consistent with the lattice transition revealed in the structural data. Except for T < 20 K (including an onset of superconductivity at $T_C$ = 1.8 K) and the vicinity of $T_S$, $\rho_a(T)$ manifests a striking linear temperature-dependence over a remarkably wide temperature range, 20 K ≤ T ≤ 700 K.

An extended regime of linear-T resistivity is a classic signature of non-Fermi liquids such as high-$T_C$ cuprates, the p-wave superconductor $Sr_2RuO_4$, Fe-based superconductors and many other correlated oxides [37], in which spin fluctuations play an important role in the electron scattering. In contrast, $Ir_3Te_8$ is diamagnetic (see below), and the application of high magnetic fields up to 14 T causes no changes the anomaly at $T_S$, or $\rho_a(T)$ at T > $T_C$ (not shown). These observations confirm the very robust linearity in $\rho_a(T)$ for $Ir_3Te_8$ must have an origin other than spin scattering. Elementary Bloch-Grüneisen theory predicts $\rho(T) \sim T^5$ for T < (0.2)$\theta_D$ ~ 49 K (the Debye temperature $\theta_D$ = 246 K for $Ir_3Te_8$), and $\rho(T) \sim T$ for T >> $\theta_D$, in the case of electron-phonon scattering. Nevertheless, few materials exhibit an extended regime of linear-T resistivity; indeed, "resistivity saturation" is anticipated when the mean-free path $l$ of the quasiparticles becomes shorter than the lattice parameter $a$ (Mott-Ioffe-Regel limit [38, 39]), or for $\rho \sim$ 100-150 $\mu\Omega$ cm (Mooij limit, [40, 41]). The experimental values of $\rho_a(T)$ for $Ir_3Te_8$ are well above the Mott-Ioffe-Regel or Mooij limits, yet $\rho_a(T)$ shows no sign of saturation up to 700 K. At low temperatures, the resistivity does not exhibit the usual $T^2$ Fermi liquid term, but remains large (≈ 0.46 m$\Omega$-cm) between $T_C$ and 10 K (see **Fig. 2a**). The striking behavior of $\rho_a(T)$ is therefore unusual and intriguing, and can be considered as "non-Fermi-liquid" behavior.

The magnetic susceptibility $\chi(T)$ is anisotropic; remarkably, $\chi_{[111]}$ is more diamagnetic than $\chi_a$, and both exhibit a first-order anomaly in the vicinity of $T_S$. An unusually strong thermal hysteresis is seen in **Fig. 2b**, consistent with a structural transition. Both $\chi_{[111]}(T)$ and $\chi_a(T)$ rapidly rise below 30 K; and a Curie-Weiss fit of $\chi_{[111]}$ (warming portion) for T < 30 K yields a Curie-Weiss temperature $\theta_{CW}$ = - 0.7 K and a small effective moment $\mu_{eff}$ = 0.024 $\mu_B$/Ir. Both the low-temperature susceptibility and the minimum in $\rho_a(T)$ near 12 K suggest a small concentration of nearly free spins is present (see the **Inset** in **Fig. 2b**).

The transition of Ir$_3$Te$_8$ to superconductivity is shown in **Fig. 3**, where both $\rho_a(T)$ and $\chi_a(T)$ exhibit onsets at $T_C$ = 1.8 K. Application of a DC magnetic field H readily depresses $T_C$ (see **Fig. 3a**), as expected. On the other hand, $\rho_a(T)$ exhibits no apparent magnetoresistive shifts for T > $T_C$, as might be anticipated in the presence of spin-flip scattering within a non-Fermi liquid. Note that $\chi_a(T)$ exhibits little difference between zero-field-cooling (ZFC) and field-cooling (FC) measurements until the temperature decreases to 0.5 K, as illustrated in **Fig. 3b**. This indicates the superconducting state may not be fully established until well below 0.5 K. Nevertheless, the Meissner effect at 0.5 K is estimated to be ~16% when the demagnetization effect is taken into account. This is a sizable superconducting volume, even if the superconducting state is incomplete at this temperature.

Our electronic structure calculations for Ir$_3$Te$_8$ were performed by first carrying out a calculation for Ir$_4$Te$_8$, then removing the Ir atoms located at the Wyckoff site *a* in a unit cell. Two bands derived from Ir-$e_g$ and Te-$5p$ orbitals cross the Fermi level $E_F$ in the case of Ir$_3$Te$_8$, and one band is quasi-two-dimensional, as shown in **Fig. 4**. The band structure of Ir$_3$Te$_8$ contrasts that of Ir$_3$Se$_8$, where only one band crosses $E_F$ [20]. The existence of a quasi-two-dimensional electronic structure in a nearly cubic lattice provides a plausible understanding of the observed lattice transition and the diamagnetism. In essence, at T > $T_S$ = 350 K, the electronic structure near the Fermi surface is formed primarily by **(1)** the $\sigma^*$ anti-bonding bonding state of two Te atoms at the center of the cubic unit cell, and **(2)** one symmetrized state formed from the $d_z^2$ orbitals of the three Ir atoms at positions (1/2,1/2,0), (0,1/2,1/2) and (1/2,0,1/2), which results in a band having strong energy dispersion along the [111] direction. At T < $T_S$, the rhombohedral phase is characterized by two distinct Wyckoff positions *a* and *e* in space group *R-3*; and the $d_z^2$ orbitals of the three Ir ions at the position *e* form a quasi-two-dimensional band. Following a general Landau-Ginzburg symmetry argument, any perturbation of the electronic structure (e.g., via even a weak electron-phonon coupling) in $Ir_3Te_8$ must lead to a lattice symmetry breaking, which explains the structural transition at $T_S$ = 350 K. When an external magnetic field is applied along the [111] direction, a diamagnetic current loop is induced to flow between the three Ir atoms coupled via the symmetrized state of the $d_z^2$ orbitals. Moreover, the expected diamagnetism must be anisotropic, and the diamagnetic response along [111] should be stronger than that along other directions, which is also consistent with the data shown in **Fig. 2b**.

Alternatively, the observed anisotropic diamagnetism can be attributed to the large SOI that breaks spin conservation. A wave-function $\Psi_\mathbf{k}$ of wave-vector $\mathbf{k}$ and spin state $|\uparrow\rangle$ or $|\downarrow\rangle$ is then of the form $\Psi_\mathbf{k} = a_\mathbf{k} |\uparrow\rangle + b_\mathbf{k} |\downarrow\rangle$, where $|a_\mathbf{k}|^2 + |b_\mathbf{k}|^2 = 1$, and m = ½ $g\mu_B H \Sigma_\mathbf{k} (|a_\mathbf{k}|^2 - |b_\mathbf{k}|^2)$. Note the net moment m can be strongly reduced with respect to the usual Pauli susceptiblitity, and can even be negative. If the standard Landau susceptibility (quantization of orbits) is added to the spin susceptibility, it is then likely to result in overall diamagnetism.

In summary, diamagnetism and linear-T resistivity of $Ir_3Te_8$ persist over an unusually wide temperature interval 20 K < T < 700 K that spans a structural transition at $T_S$ = 350 K. However, there is no clear evidence for strong spin fluctuations that usually underpin non-Fermi liquid effects. The observed lattice transition and diamagnetism could be attributed to a quasi-two-dimensional band crossing at the Fermi level. However, the relationship of the superconducting state below $T_C$ = 1.8 K to the unusual collection of normal state properties of $Ir_3Te_8$ poses a challenge to our understanding of heavy transition metal materials.

This work was supported by NSF through grants DMR-0856234 and EPS-0814194; JPH acknowledges support by the Ministry of Science and Technology of China 973 program (2012CB821400) and NSFC-1190024. LED and PS acknowledge the support of U.S. Dept. of Energy Grants No. DE-FG02-97ER45653 and DE-FG02-98ER45707, respectively. XXW thanks H. M. Weng for enlightening discussions.
# High school student, Paul Lawrence Dunbar High School, Lexington, KY 40513

**Captions**

**Fig.1. (a)** The crystal structure of $Ir_3Te_8$; **(b)** the definition of the bond angles, $\theta_1$ and $\theta_2$ (upper panel) and the temperature dependence of $\theta_1$ and $\theta_2$ (lower panel); **(c)** a representative x-ray diffraction pattern at T=295 K for [$h$ 1 $l$]; and **(d)** representative single crystals of $Ir_3Te_8$.

**Fig. 2.** The temperature dependence of **(a)** the a-axis resistivity $\rho_a(T)$ for $1.7\ K \leq T \leq 700\ K$; and **(b)** the magnetic susceptibility $\chi_{[111]}(T)$ along the direction [111] and $\chi_a(T)$ along the a-axis at $\mu_oH = 0.5$ T for $1.7\ K \leq T \leq 395\ K$. **Inset**: $(\Delta\chi)^{-1}_{[111]}$ vs. T for $1.7\ K \leq T \leq 30\ K$, where $\Delta\chi_{[111]} \equiv \chi_{[111]} - \chi_o$, and $\chi_o = -0.000372$ emu/mole).

**Fig. 3.** The temperature dependence of **(a)** the a-axis resistivity $\rho_a(T)$ for $1.7\ K \leq T \leq 6\ K$ at $\mu_oH = 0, 0.5, 1$, and 5 T; and **(b)** the a-axis magnetic susceptibility $\chi_a(T)$ for $0.5\ K \leq T \leq 6\ K$ at $\mu_oH = 0.005$ T.

**Fig.4.** The band structures and density of states (DOS) of **(a)** $Ir_4Te_8$ and **(b)** $Ir_3Te_8$; and **(c)** Fermi surfaces for $Ir_3Te_8$: the lower band (left) and the higher band (right). $\Gamma$ is the center of the Brillouin Zone, R (0.5, 0.5, 0.5), X (0, 0.5, 0) and M (0.5, 0.5, 0). The total DOS and momentum projected-DOS are shown in the middle panel and right panel in **(a)** and **(b)**, respectively. The middle letters are the Wyckoff positions; for instance, "Ir-e-5d" represents the 5d orbital projected-DOS of Ir at positions *e*.

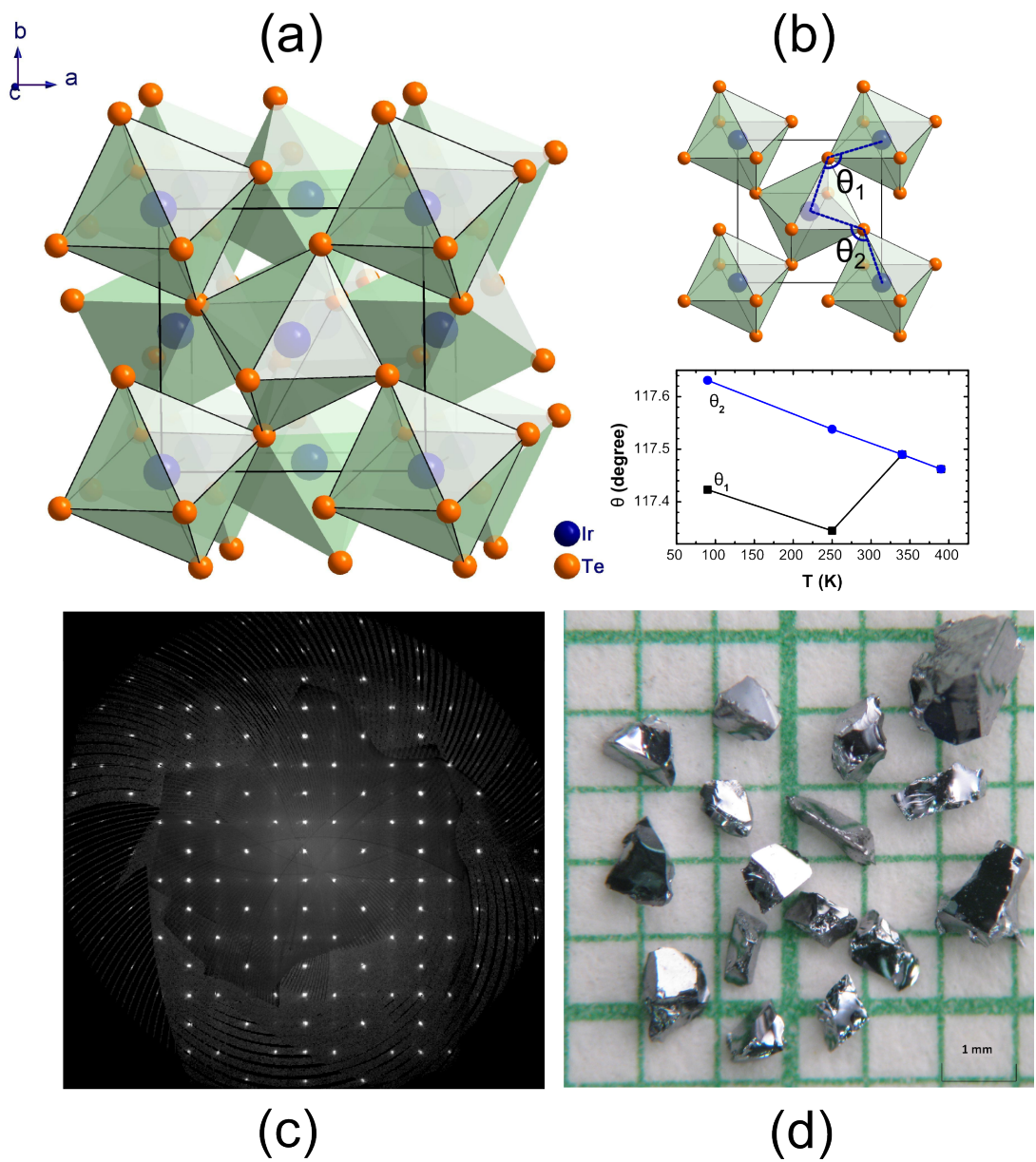

Fig.1

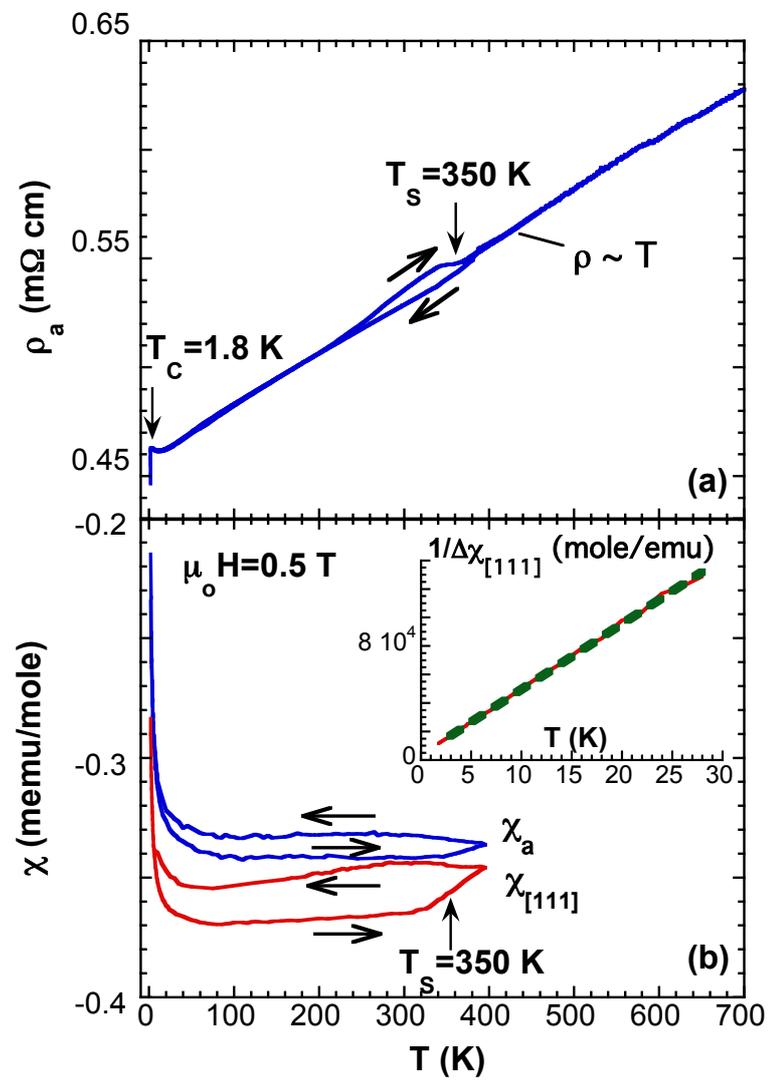

**Fig. 2**

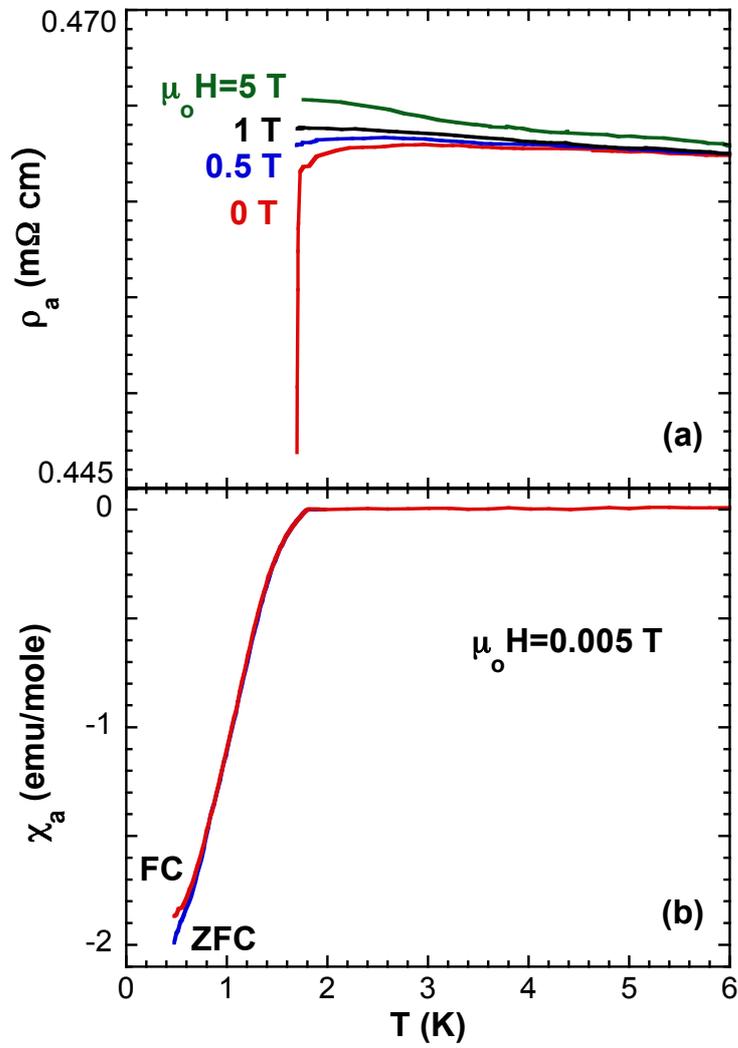

Fig. 3

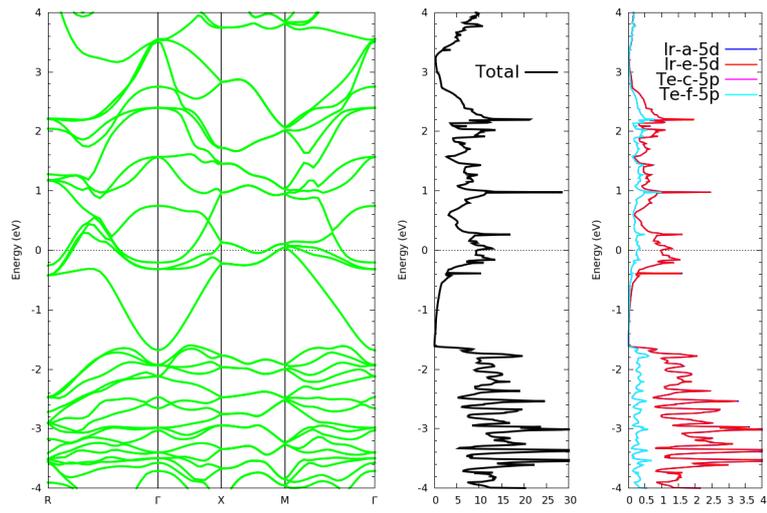

(a)

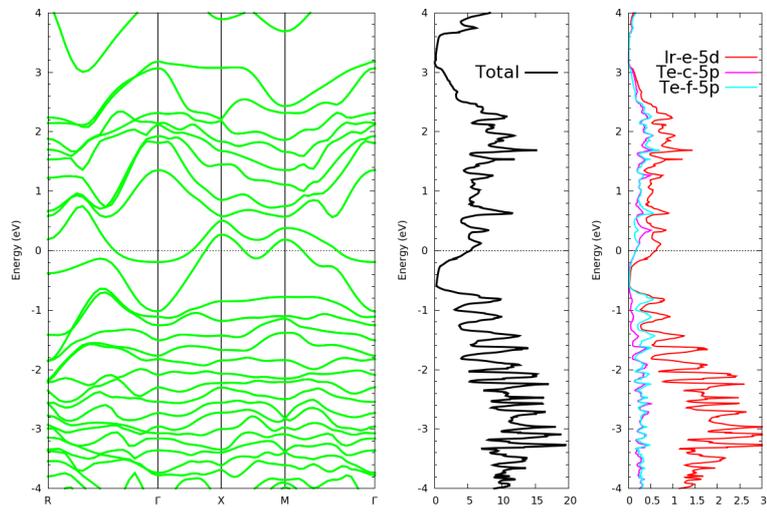

(b)

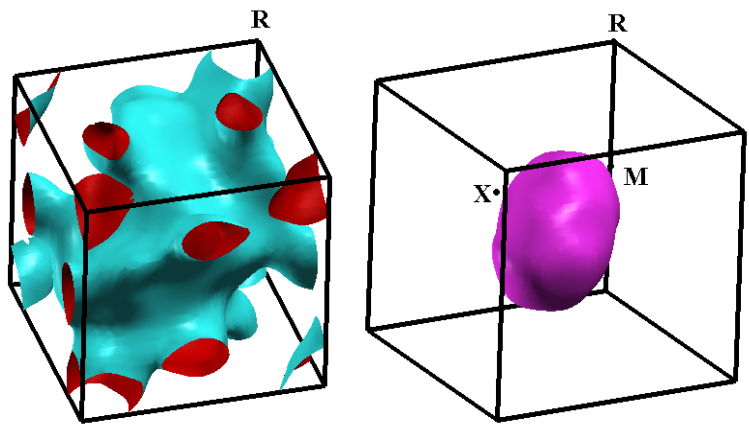

(c)    **Fig. 4**

17